\documentclass[12pt, a4paper, oneside]{article}
\usepackage{srcltx,amsmath}
\usepackage{amssymb}
\usepackage[utf8]{inputenc}
\usepackage{hyperref}
\usepackage{simplewick}

\begin{document}

\title{Pieri rules, vertex operators  and Baxter Q-matrix
\footnote{In honour of Rodney Baxter's 75th birthday.}}
\author{Antoine Duval and Vincent Pasquier\\
Institut de Physique Théorique, DSM, CEA, URA2306 CNRS\\
Saclay, F-91191 Gif-sur-Yvette, France
}

\maketitle
\begin{abstract}
We use the Pieri rules to recover the q-boson model and show it is equivalent to a discretized version of the relativistic Toda chain. 
We identify its semi infinite transfer matrix and the corresponding Baxter Q-matrix 
with half vertex operators related by an $\omega$-duality transformation.
We observe that the scalar product of two higher spin XXZ wave functions can be expressed with a Gaudin determinant.

\end{abstract}

\section{ Introduction}

In this paper, we establish a correspondence between some lattice integrable models, 
the q-bosons and the Toda lattice, and multi-variable symmetric  orthogonal polynomials.
Our starting  observation is that the Pieri rules \cite{macdonald_symmetric_1999}
provide 
a family of commuting transfer matrices which are the main tool to study lattice integrable models.
They are the multi-variable generalization of the three terms recursion relations of orthogonal polynomials \cite{szego_orthogonal_1939}.
Given a family of elementary symmetric functions organized into a generating function $\Omega(u)$,
and a basis of  symmetric orthogonal polynomials $P_\lambda$, they tell us how the product $\Omega(u)P_\mu$ decomposes onto
the $P_\lambda$-basis. The branching rule of this decomposition $\Omega(u)P_\mu=\sum_{\lambda}P_{\lambda}T_{\lambda\mu}(u)$
define a commuting family of transfer matrices and can be interpreted as the eigenvalue equation
for  $T(u)$.

In favorable cases, the transfer matrix is a product of local Lax operators ($L$ matrices \cite{faddeev_how_1996}) and one 
can deduce from it the local Boltzmann weights of a lattice model.
The lattice model we analyze in this respect is the q-boson model \cite{kulish_q_1990} where Bosonic particles occupy the sites of
a one dimensional chain. 
The dynamics is controlled by a transfer matrix which at each transfer time lets the particles  simultaneously hop
in one direction in between their position and the next occupied one with an
intensity controlled by the deformation parameter of the q-bosons. 
When the chain is closed, the number of particles  is conserved, 
and it can be diagonalized by coordinate
Bethe ansatz. The components of its eigenvectors are  Hall-Littlewood polynomials $P_\mu$ evaluated at
roots of the Bethe equations.

We study this model on a
semi infinite chain where there is no quantization condition (Bethe equations).
We introduce two different dynamics. In one, the particles simultaneously
hop to their right with possible creation at the origin. In the other, they hop to their left with possible annihilation at the origin.
The corresponding transfer matrices  are obtained as the limit of finite size matrices
whose commutations relations are dictated by the Yang-Baxter equation. In the limit,  these commutations simplify and 
can be represented with simple  Vertex operators as was discovered by by Tsilevich \cite{tsilevich_quantum_2005}.
When the deformation parameter is equal to zero, using the Frobenius correspondence 
between partitions and Fock space configurations,
one recovers a free fermion model.
The components of the eigenvectors are
the Hall-Littlewood polynomials obeying a Pieri dynamics as described above, generated 
by the eigenvalue of the Bethe equations.

A remarkable property of the Hall-Littlewood polynomials is the existence of a second Pieri rule (related to the Hall algebra). 
We repeat the same analyses for its dynamics, relabeling
the configurations in terms of a discretized version of the Toda lattice. Unlike the q-bosons case,
the Hall-Pieri transfer matrix is not of a Lax type but is instead the Baxter Q-matrix \cite{rodney_baxter_exactly_2008} generating the Bäcklund 
Toda lattice dynamics \cite{sklyanin_backlund_2000}. We construct it as an alternative to the coordinate Bethe ansatz, 
to derive the Toda chain Bethe equations. We use its open variant  to recover  the open Toda chain wave functions
in the same way as Gerasimov Lebedev and Oblezin \cite{gerasimov_q-deformed_2009}. We identify it with a vertex operator, the matrix
elements of which are the branching coefficients of
the Hall-Pieri rules and we recover this relation as a matrix element of the Yang-Baxter equation. 

The open Toda chain wave functions can be deduced by an  $\omega$-duality transformation from
the Hall-Littlewood polynomials.  We have extensively used the $\lambda$-ring formalism 
\cite{lascoux_symmetric_????}\cite{garsia_adriano_explicit_1999} unifying the two type
of Pieri rules and enabling to relate them simply to Cauchy identities.

The q-boson model appears as the infinite spin limit of an XXZ (or six vertex) higher spin chain which can be 
diagonalized by coordinate Bethe
ansatz. In the last section, we observe that the scalar product of two semi infinite chain
wave functions is independent of the spin and can be expressed with a Gaudin determinant \cite{gaudin_michel_bose_1971}.

\section{The q-bosons \label{q-boson}}

The q-bosons \cite{kulish_q_1990}\cite{bogoliubov_correlation_1998}\cite{van_diejen_diagonalization_2014} are defined
by their commutation relations:
\begin{eqnarray}
tS\bar S-\bar S S=t-1,
\label{SS}  
\end{eqnarray}
and are realized with a Weyl pair $\tau,S,\ \tau S=tS\tau$ as:
\begin{eqnarray}
\bar S=S^{-1}(1-\tau).
\label{SS1} 
\end{eqnarray}


There are two possible Hermitian conjugations which preserve the relations (\ref{SS})
depending on the value of the parameter $t$, $t^*=t$
or $t^*=t^{-1}$. We shall only consider the first
one here $S^\dagger=\bar S$, with $t$ a real number between $0$ and $1$.

$S$ and $\bar S$ act on q-bosons states which can be identified with $S$ polynomials, $S^n=|{n}\rangle$ 
($\bar S$ equal to $0$ when sitting to the right of  $S^n$) with  $n$ a nonnegative integer
as $\bar S|{n}\rangle=(1-t^{n})|{n_{}-1}\rangle$, $\tau|{n}\rangle=t^{n}|{n_{}}\rangle$.
The scalar product such that $\bar S$ is the adjoint of $S$, ($\bar \tau=\tau$) is:
\begin{eqnarray} 
\langle {n}|{m}\rangle&=& (t)_m\delta_{n,m},
\label{scal}
\end{eqnarray}
with:
 \begin{eqnarray} 
(a)_m=(1-a)\cdots(1-at^{m-1}).
\label{blanda}
\end{eqnarray}

 The configurations space of the q-boson spin chain of length $N$ consists in states 
 where sites labeled by  $k$ are occupied by $m_k$ q-bosons, $S_0^{m_{0}}S_1^{m_{1}}\cdots S_{N-1}^{m_{N-1}}$ 
 with $\sum_k m_k=n$, the total number of Bosons.
 We can also characterize the state by the sequence of positions of the Bosons ordered decreasingly: $(N-1\ge\mu_{1}\ge\cdots\ge\mu_n\ge 0)$, 
 $|(N-1)^{m_{N-1}}\cdots 0^{m_{0}} \rangle=|\mu_n\cdots\mu_1\rangle$ where the particles labeled $i\ (1\le i\le n)$ forgetting about their Bosonic nature
 occupies the sites $\mu_i,\ (0\le\mu_i\le N-1)$  of the chain.
 The scalar product between two Boson states is then given by:
\begin{eqnarray} 
\langle \mu|\nu\rangle&=& \delta_{\mu,\nu} \prod_0^{N-1} {m_{k}!_t}.
\label{scal1}
\end{eqnarray}
where $(n)!_t=(t)_n$ is the t-factorial.

We shall also consider the trivial representation $S_0=\bar S_0=1$, in which case,
the occupation number starts at $k=1,\ \mu_i\ge1$. Equivalently, the site
$0$ is occupied by an infinite number of bosons $m_0=\infty$.

 We consider the Lax-Matrix:
 \begin{eqnarray}
L_k(z)=
\left( \begin{array}{cc}
1&  z\bar S_k\\S_k& z
\end{array}\right)
\label{laxq1} 
\end{eqnarray}
where the q-boson creation and annihilation operators $S_k,\ \bar S_k$ act independently
on spaces labeled $k$.
It obeys the Yang-Baxter equation \cite{faddeev_how_1996}:
\begin{eqnarray}
 R_{12} L^1(u)L^2(v)= L^2(v)L^1(u)R_{12},
\label{yb}
\end{eqnarray}
with $L^1=L\otimes 1,\ L^2=1\otimes L$.
and the $R$ matrix acting in the tensor product is:
\begin{eqnarray}
R_{12}(u,v)=\left( \begin{array}{cccc}
a&.&.&. \\.&b& \bar c&\\.&c& \bar b&.\\.&.&.&a
\end{array}\right)
\label{R}
\end{eqnarray}
\begin{eqnarray}
a &=&ut-v,\ b =u-v,\ \bar b =t(u-v),\nonumber\\
c&=&v(t-1),\ \bar c =u(t-1).
\end{eqnarray}
Everywhere, $.=0$ in the following.
It is straightforward to verify that the relations (\ref{R}) with $L$ given by (\ref{laxq1}) are equivalent
to the q-boson relations (\ref{SS}) satisfied by $S_i,\bar S_i$ which commute for different values of $i$.
 
 The monodromy matrix:
 \begin{eqnarray} 
T_N(z)
=L_0L_1\cdots L_{N-1}
\label{monodromy}
\end{eqnarray}
takes the form:
 \begin{eqnarray} 
T_N(z)=\pi
\left( \begin{array}{c}
 1 \\S_0
\end{array}\right)      
 \prod_1^{N-1} (1+z\bar S_{i-1}S_i)
\left( \begin{array}{cc}
  1&z\bar S_{N-1}
\end{array}\right) ,
\label{Tp0}
\end{eqnarray}
where the projector $\pi$ requires we erase $S_i$ and $\bar S_i$ in the terms of the $z$ expansion each time they occur in pairs.
In other words, $\pi$ means we set $S_i\bar S_i=1$ in the formal expansion of the above product but do not identify $\bar S_i$ with $S_i^{-1}$.
In particular, the transfer Matrix of the periodic chain is:
 \begin{eqnarray} 
\Lambda_{N,x}(z)={\rm  Tr}\ T_N(z)D
\label{perioddef}
\end{eqnarray}
where $D$ is the matrix:
 \begin{eqnarray} 
D=\left( \begin{array}{cc}
1&  .\\.& x
\end{array}\right)
\label{Ydef}
\end{eqnarray}
define a commuting family of transfer matrices.
It takes the form:
 \begin{eqnarray} 
\Lambda_{N,x}(z)=\pi    (1+z\bar S_0S_1)\cdots (1+z\bar S_{N-1}S_{N}),
\label{TraceT}
\end{eqnarray}
where we require the periodicity condition:
\begin{eqnarray} 
S_{N+k}=xS_k.
\label{periodicite}
\end{eqnarray} 
The configuration space consists of
decreasing sequences  $\lambda$ ($1\le \lambda_i\le N$), and
the matrix elements $\Lambda_{\lambda\mu}$ can be nonzero only when each boson jumps to the right in between his position and the next 
occupied one: $\lambda_{n+1}\le \mu_n\leq \lambda_{n}\leq \mu_{n-1}\cdots\leq \lambda_1=\lambda_{n+1}+N$.

The one step translation $\mathcal T$: 
\begin{eqnarray} 
\mathcal TS_k=S_{k+1}\mathcal T
\label{translate}
\end{eqnarray} 
commutes with the transfer matrix: $\mathcal T\Lambda_{N,x}=\Lambda_{N,x}\mathcal T$,
and in the $n$-Boson sector $\mathcal T^N=x^n$. 

Everywhere, we use the symbol $\check u$ to represent  the inverse of a variable: $\check u=1/u$. 
We require:
 \begin{eqnarray} 
\bar x&=&\check x,\nonumber\\ \bar z&=&\check z.
\label{baryz}
\end{eqnarray}
We can rewrite
$\pi\cdots(1+z\bar S_iS_{i+1})$ as $\pi\cdots z\bar S_{i}(1+\check zS_i\bar S_{i+1})S_{i+1}$, 
so that the Hermitian property $\bar \Lambda_{N,x}( z)=\check x \check z^N \Lambda_{N,x}(z)$ holds.
In other words, $ \Lambda^{\rm H}= \bar \Lambda_{N,X}( z)/\sqrt{xz^N}$ is a Hermitian operator:
\begin{eqnarray} 
\bar \Lambda^{\rm H}= \Lambda^{\rm H}.
\label{hermiteT}
\end{eqnarray}

This transfer matrix can be diagonalized by the coordinate Bethe ansatz.
We obtain the eigenvectors and corresponding eigenvalues in the appendix \ref{lieb}
as a limit of the spin XXZ chain Bethe equations (\ref{lax-sl2}).

\section{Finite and semi infinite open chains\label{half}}

Here, we  consider the q-boson model on an open chain satisfying some specific boundary conditions.
Our aim is to let the size of this chain go to infinity and relate it to some vertex operator constructions 
\cite{okounkov_correlation_2003}\cite{betea_perfect_2014}. The material of this section was essentially discovered
in the precursor work of Tsilevich \cite{tsilevich_quantum_2005}, and
overlaps with similar constructions  \cite{korff_cylindric_2012}, \cite{borodin_spectral_2015}.
More specifically, the relation between the semi-infinite chain and  vertex operators discussed here 
appears in the recent preprint of Wheeler and Zinn-Justin \cite{wheeler_refined_2015}.
We follow the approach of Macdonald \cite{macdonald_symmetric_1999},
to emphasize the connections between the symmetric polynomials and the semi-infinite transfer matrices we introduce
and to familiarize the reader with the $\lambda$-ring notations.

We take the trivial representation $S_0=\bar S_0=1$ for $L_0$
and the q-boson representation for $L_i,\ i>0$, so $T_{11}=T_{21},\ T_{12}=T_{22}$.
There is no occupation on site $0$ and the chain starts at site $1$. The configuration space consists of
partitions $\lambda$ ($1\le \lambda_i\le N$). 
The matrix elements $(T_{11})_{\lambda\mu}$ can be nonzero only when each boson jumps to the right in between his position and the next 
occupied one, and an additional boson labeled $n+1$ can be injected before the first occupied site: 
$0\le \lambda_{n+1}\le \mu_n\leq \lambda_{n}\leq \mu_{n-1}\cdots\leq \lambda_1\le N$ 
(No boson is injected if $\lambda_{n+1}=0$, and the number of Bosons $n$  increases by one unit otherwise).
We denote this condition $\mu\prec \lambda$,
or $\lambda-\mu$ is a horizontal strip. Similarly, the matrix elements of $T_{12}$ can 
be nonzero only when each boson jumps to the left in between his position and the preceding
occupied one, and the leftmost boson (at position $\mu_n$) can disappear. 

We define $A_N$ and its hermitian conjugated $\bar A_N$ by:
 \begin{eqnarray} 
(A_N, z^{N+1}\bar A_N)=((T_{N+1})_{11},((T_{N+1})_{12})=(1,z)L_1L_2\cdots L_{N}
\label{ABinfini}
\end{eqnarray}

Explicitly,
\begin{eqnarray} 
A_N(z)&=&\pi(1+zS_1)(1+zS_2\bar S_{1})\cdots (1+zS_N\bar S_{N-1})\nonumber\\
\bar A_N(z)&=&\pi(1+\check z^{}\bar S_1)(1+\check z^{}\bar S_2 S_{1})\cdots (1+\check z^{}\bar S_N S_{N-1}),
\label{gamma-infini}
\end{eqnarray}

Some commutation relations which follow from the $RLL$ relation (\ref{yb}) are:
\begin{eqnarray} 
[T_{11}(u),T_{11}(v)]&=&[T_{12}(u),T_{12}(v)]=0\nonumber\\
a T_{11}(u)T_{12}(v)&=&bT_{12}(v)T_{11}(u)+cT_{11}(v)T_{12}(u)
\label{ABcommut}
\end{eqnarray}
independently of $N$. 

We consider a semi infinite chain obtained by taking the $N\to \infty$  limit of the monodromy matrix
acting in a space with a finite number of q-bosons.
In the limit $N\to \infty$ of the finite $N$ Hilbert space, 
we define the matrices (which coincide with the fermion limit of \cite{betea_perfect_2014} for $t=0$):
 \begin{eqnarray} 
(\Gamma_-,\Gamma_+)(z)=\lim_{N\to \infty} (A_N,\bar A_N).
\label{Gammadef}
\end{eqnarray}

$\Gamma_-$  is defined for  $|z|<1$ and coincides with the limit of $A_N(z)$; $\Gamma_+$ for  $|z|>1$
coincides with the limit of $\bar A_N(z)$. 
They obey the same hermitian property as $A_N,\bar A_N$:
\begin{eqnarray} 
\bar\Gamma_-(z)=\Gamma_+ (z).
\label{Gamma-conjug}
\end{eqnarray}
Expressed in terms of $A,\bar A$, the second line of (\ref{ABcommut}) rewrites:
\begin{eqnarray} 
a A_N(u)\bar A_N(v)=b\bar A_N(v)A_N(u)+c(u/v)^{N+1}A_N(v)\bar A_N(u).
\label{ABcommut1}
\end{eqnarray}
In the $N\to \infty$ limit, the third term behaves as $|u/v|^N<<1$
compared to the two others. So, the commutations of $\Gamma_\pm$ in the $N\to \infty$ limit are:
\begin{eqnarray} 
[\Gamma_-(v),\Gamma_-(v')]&=&[\Gamma_+(u),\Gamma_+(u')]=0\nonumber\\
\Gamma_+(u)\Gamma_-(v)&=&{1-tv/u\over 1-v/u}\Gamma_-(v)\Gamma_+(u)
\label{gammainfini}
\end{eqnarray}
where $|u|,|u'|>1$ and  $|v|,|v'|<1$.

$\Gamma_+,\Gamma_-$ respectively act as the identity on the left and right  vacuum $\langle .|$ annihilated by $S_i$
and $|.\rangle$ annihilated by $\bar S_i$.

The  expression of the matrix elements of $\Gamma_-(z)$ are  given by:
\begin{eqnarray}
\Gamma_-(z)|\mu\rangle
=\sum_{\mu\prec \lambda}z^{|\lambda-\mu|}\psi_{\lambda/\mu}|\lambda\rangle
\label{gauche1}
\end{eqnarray}
where:
\begin{eqnarray}
\psi_{\lambda/\mu}=\prod_{m_j(\lambda)=m_j(\mu)-1}(1-t^{m_j(\mu)})
\label{psidef}
\end{eqnarray}
agrees with the definition  \cite{macdonald_symmetric_1999} (5.8') p.229.

Similarly, the matrix elements of $\Gamma_+(z)$ are  given by:
\begin{eqnarray}
\Gamma_+(z)|\lambda\rangle
=\sum_{\mu\prec \lambda}\check z^{|\lambda-\mu|}\phi_{\lambda/\mu}|\mu\rangle
\label{gauche11}
\end{eqnarray}
where:
\begin{eqnarray}
\phi_{\lambda/\mu}=\prod_{m_j(\lambda)=m_j(\mu)+1}(1-t^{m_j(\lambda)})
\label{phidef}
\end{eqnarray}
coincides with the definition  \cite{macdonald_symmetric_1999} (5.8) p.228.

Define the state:
\begin{eqnarray} 
 \langle  {U}|=\langle .|\Gamma_+( u_1)\Gamma_+( u_2)\cdots\Gamma_+( u_n)=\langle .|\Gamma_+( U)
\label{gauche}
\end{eqnarray}
where $ U=\{u_k,\ 1\le k\le n\}=u_1+\cdots+u_n$, meaning that
the order of the $u_i$ is irrelevant
(due to the commutations of $\Gamma_+( u_i)$).
More generally,  we denote $\langle U+  V|=\langle  U|\Gamma_+( V)$.

We can use the commutation relations (\ref{gammainfini}) to check that  $\langle U|$
is a left eigenstate of  $\Gamma_-(z)$.
Using $\lambda$-ring notations \cite{lascoux_symmetric_????} \cite{garsia_adriano_explicit_1999}:
\begin{eqnarray}
\langle U|\Gamma_-(z)=\Omega_t(\check Uz) \langle U|
\label{gauche2}
\end{eqnarray}
where $ \check U=\check u_1+\cdots+\check u_n$, and
$\Omega(X)$ is the generating function of the complete symmetric functions \cite{macdonald_symmetric_1999} (2.5) p 21:
\begin{eqnarray}
\Omega(X)=\prod_{x_k\in X}{1\over 1-x_k}
\label{Hsymdef}
\end{eqnarray}
We also use $\lambda$-ring notations to define  $\Omega_t(X)=\Omega(X(1-t))$. Namely
since $\Omega(X\pm Y)=\Omega(X)\Omega(Y)^{\pm 1}$, $\Omega_t(X)=\Omega(X)/\Omega(tX)$.

Putting (\ref{gauche1}) and (\ref{gauche2}) together, we obtain the eigenvalue equation:
\begin{eqnarray}
\Omega_t(\check Uz) \langle U|\mu\rangle
=\sum_{\mu\prec \lambda}z^{|\lambda-\mu|}\psi_{\lambda/\mu}\langle U|\lambda\rangle.
\label{gauche2ensemble}
\end{eqnarray}
Let  $\Omega_t(z\check U)=\sum_{r=0}^\infty q_r(\check U)z^r$
where $q_r$ defines elementary symmetric polynomials of degree $r$. 
Define the polynomials $Q_\mu(\check U)$ symmetric in $\check U$ (The symmetry property
follows from the commutations of $\Gamma_+( U)$), homogeneous of degree $|\mu|$ from the expansion:
\begin{eqnarray} 
\langle U|=
\sum_\lambda 
Q_{\lambda} (\check U)
\langle\bar \lambda|
\label{Pieri2}
\end{eqnarray}
where $\langle \bar \lambda|=\langle \lambda|/\langle \lambda|\lambda\rangle$ is dual to $|\lambda\rangle$. 
Since  the partitions $\lambda$'s which occur in (\ref{Pieri2}) all have their length $l(\lambda)\ge |U|$, the polynomial $Q_{\lambda} (\check U)$
is null if the number of variables $u_i$ is less than  $l(\lambda)$. On the other hand, since $\Gamma_-(0)=1$, 
the number of variables in $Q_\lambda$ is immaterial
as long as it is larger or equal to the length of $\lambda$ and can therefore be sent to infinity.

Expanding (\ref{gauche2ensemble}) in $z$, yields the decomposition of 
$q_rQ_\mu$:
\begin{eqnarray} 
q_rQ_\mu=\sum_{\mu\prec \lambda, |\lambda-\mu|=r} \psi_{\lambda/\mu} Q_\lambda
\label{pieri}
\end{eqnarray}
with $\lambda-\mu$ a horizontal r-strip.
It coincides with \cite{macdonald_symmetric_1999} (5.7') p.229,
and allows to identify $Q_\mu$ with the Hall-Littlewood polynomials.

Acting with $\Gamma_+(U')$ on a state $\langle\bar U|$ (instead of  $\langle .|$ as in (\ref{gauche})
enables to define another symmetric function $Q_{\lambda/\mu}(\check U)$ for $\mu\in\lambda$:
\begin{eqnarray} 
Q_{\lambda}(\check U+\check U')= \sum_{\mu\subset\lambda} Q_{\lambda/\mu} (\check U)Q_{\mu} (\check U').
\label{produitdeq}
\end{eqnarray}
Equivalently:
\begin{eqnarray} 
\langle \bar\mu|\Gamma_+( U)=
\sum_ {\mu\subset\lambda}
Q_{\lambda/\mu} (\check U)
\langle\bar \lambda|
\label{qla/mu}.
\end{eqnarray}

By taking the adjoint of (\ref{gauche}), we define the state:
\begin{eqnarray} 
|{V}\rangle=\Gamma_-(v_1)\Gamma_-(v_2)\cdots\Gamma_-(v_n)|.\rangle=\Gamma_-(V)|.\rangle
 =\sum_\lambda P_\lambda (V) |\lambda \rangle
\label{droit}
\end{eqnarray}
where
$P_{\lambda}=Q_{\lambda}/\langle \lambda|\lambda\rangle$ and $V=\{v_k,\ 1\le k\le m\}$.
It is a right eigenvector of $\Gamma_+( z)$:
\begin{eqnarray}
 \Gamma_+( z)|{ V}\rangle =\Omega_t(\check z V)|{ V}\rangle.
\label{droit2}
\end{eqnarray}

More generally,
\begin{eqnarray} 
\Gamma_-(v_1)\Gamma_-(v_2)\cdots\Gamma_-(v_n)|\mu\rangle=\Gamma_-(V)|\mu\rangle
 =\sum_\lambda P_{\lambda/\mu} (V) |\lambda \rangle
\label{droitt}
\end{eqnarray}
If we introduce the sequences of partitions:
$\lambda_n=\mu\prec \lambda_{n-1}\cdots \prec \lambda_0=\lambda$
which form the tableaux $T$ of shape $\lambda-\mu$ \cite{macdonald_symmetric_1999} p.5,
and the quantities:
\begin{eqnarray} 
v^T=v_1^{|\lambda_0-\lambda_1|}\cdots v_n^{|\lambda_{n-1}-\lambda_n|}\nonumber\\
\psi_T=\psi_{\lambda_0/\lambda_1}\cdots\psi_{\lambda_{n-1}/\lambda_n}.
\label{droittdef1}
\end{eqnarray}
Rewriting (\ref{droitt}) in matrix form, we obtain the tableau expression of $P_{\lambda/\mu}$
\cite{macdonald_symmetric_1999} 5.11' p.229:
\begin{eqnarray} 
P_{\lambda/\mu}(V)=\sum_T \psi_T v^T
\label{tableau}
\end{eqnarray}
where the summation is over all the tableaux $T$ of shape $\lambda-\mu$.

Inserting the identity $1=\sum_\lambda |\lambda\rangle \langle \bar \lambda|$ into the scalar product
$\langle U|V\rangle$ yields the Cauchy identity:
\begin{eqnarray} 
\langle U|V\rangle=\sum_\lambda  Q_\lambda(\check{U})P_\lambda(V)
=\Omega_t(\check{U}V).
\label{cauchy}
\end{eqnarray}

The product $P_\mu P_\nu$ can be decomposed on the $P_\lambda$ basis:
\begin{eqnarray} 
P_\mu P_\nu=\sum_\lambda f^\lambda_{\mu\nu} P_\lambda.
\label{produit}
\end{eqnarray}
We have:
\begin{eqnarray} 
\langle\bar\mu|\Gamma_+( U)|V\rangle=Q_{\lambda/\mu}(\check U)P_\lambda( V)=\Omega_t(\check{U}V)P_\mu( V).
\label{deuxfacons}
\end{eqnarray}
Using the Cauchy identity to decompose $\Omega_t(\check{U}V)$ and identifying the coefficient of $P_\lambda$
in the resulting expression, we recover the definition of $Q_{\lambda/\mu}$ \cite{macdonald_symmetric_1999} (5.2) p.227:
\begin{eqnarray} 
Q_{\lambda/\mu}=\sum_\nu f^\lambda_{\mu\nu}Q_\nu.
\label{ql/r}
\end{eqnarray}

We can obtain a more explicit expression of $Q_\mu$ by diagonalizing
the infinite chain by Bethe Anzats \cite{borodin_spectral_2015} in the sector of $n$ q-bosons.
The infinite chain  transfer matrix is defined as:
 \begin{eqnarray} 
A(z)&=&\pi\prod_{-\infty}^{-\infty}(1+zS_{k+1}\bar S_{k}),
\label{Tpinfini}
\end{eqnarray}
in which case the sites take integer values $\mu_k\in\mathbb{Z}$ defining  a sequence of decreasing integers, $\mu_n\ge\cdots\ge\mu_1$.

The transfer matrix elements have the same form as (\ref{gauche1})
with no positivity restriction on $\mu_i$:
\begin{eqnarray}
A(z)|\mu\rangle
=\sum_{\mu\prec \lambda}z^{|\lambda-\mu|}\psi_{\lambda/\mu}|\lambda\rangle
\label{Tgauche1}
\end{eqnarray}
and $\lambda,\mu$ have length $n$.
The Bethe-Lieb left eigenvectors, $\langle R|\mu\rangle=R_\mu$, corresponding to the eigenvalue (\ref{gauche2}):
\begin{eqnarray}
\sum_\lambda R_\lambda(\check U)A_{\lambda\mu}(z)=\Omega_t(\check U z)R_\mu(\check U)
\label{betheeigen}
\end{eqnarray}
has the expression (appendix (\ref{lieb}):
\begin{eqnarray}
R_\mu(\check U)=\sum_P(\check  u_1^{\mu_1}\cdots \check u_n^{\mu_n}\prod_{i<j} {\check u_i-t\check u_j\over\check  u_i-\check u_j})
\label{bethe}
\end{eqnarray}
summed over the permutations of the $\check u_i$. One recognizes the polynomial
$R_\mu(\check U)$ p. 204 of \cite{macdonald_symmetric_1999}, where we  extend its definition to sequences of decreasing integers.

Consider the restriction of (\ref{Tgauche1}) 
to configurations  of $n$ q-bosons sitting on non-negative sites,
$|0^{m_0}\mu\rangle$ where $\mu$ is a partition, and  $m_0=n-l_\mu$. Since $(1+zS_1)$ in $\Gamma_-(z)$ (\ref{Tp0}) is replaced by $(1+zS_1\bar S_0)$ in
$T(z)$, they are related by a similarity transform: $T_{\lambda\mu}=(t)_{n-l_\mu}/( t)_{n-l_\lambda} (\Gamma_-)_{\lambda\mu}$ 
when restricted to these configurations. 
So, the eigenvectors of $T(z)$ are deduced from those of $\Gamma_-(z)$ by:
\begin{eqnarray} 
{(1-t)^n\over ( t)_{m_0}}R_{0^{m_0}\mu}(\check U)=Q_\mu(\check U)
\label{rtoq}
\end{eqnarray} 
where the normalization $(1-t)^n$ is required for $Q_\mu$ not to depend on $m_0$.
In this way, we recover the expression of $Q_\mu(\check U)$ \cite{macdonald_symmetric_1999} p 211,
and  $P_\mu(V)$ p 208.

\section{The half vertex-operators $\Gamma_{a,\pm}$ \label{halfa}}

In this section, we use more extensively the $\lambda$-ring formalism to introduce the  vertex operators $\omega$-dual
to those of the preceding section. We introduce another Pieri rule and  use its branching coefficients to construct
a transfer matrix  which we shall derive in a different way from its commutation relations in the section (\ref{glomac1}).

The scalar product analogous to (\ref{scal1})
on symmetric polynomials:
\begin{eqnarray} 
\langle Q_\mu|Q_\nu\rangle=\delta_{\mu,\nu} \prod_{k\ge 1}m_{k}!_t
\label{scal10}
\end{eqnarray}
\cite{macdonald_symmetric_1999} (4.9) p.225
enables to formally identify the dual states $\langle \bar\lambda|,\ |\lambda \rangle$ with the dual polynomials $P_\lambda( X),\ Q_\lambda( X)$ in an
infinite number of variables $|X|=\infty$,
so that using the Cauchy relation, (\ref{droit}) rewrites:
\begin{eqnarray} 
|{V}\rangle&=&\Omega_t(VX)
\label{nouvX}
\end{eqnarray}
and the transfer matrix $\Gamma_-(z)$ act multiplicatively as:
\begin{eqnarray} 
\Gamma_-(z)\alpha(X)=\Omega_t(zX)\alpha(X).
\label{multip}
\end{eqnarray}

The scalar product can also be defined by its reproducing Kernel \cite{garsia_adriano_explicit_1999}:
\begin{eqnarray} 
\langle \alpha(X)|\Omega_t(XY)\rangle=\alpha(Y).
\label{scal101}
\end{eqnarray}
$\Gamma_+(z)$is defined by the Hermitian condition (\ref{Gamma-conjug}) $\Gamma_+(z)=\bar\Gamma_- ( z)$, and from
$
   \langle\Gamma_-(z) \alpha|\Omega_t(XY)\rangle=\langle\Omega_t(zX) \alpha|\Omega_t(XY)\rangle
   =\langle \alpha|\Omega_t((X+z)Y)\rangle
$
we deduce $\Gamma_+(z)$ acts by ``adding $\check z$ to $X$'' \cite{garsia_adriano_explicit_1999}\cite{lascoux_adding_2007}:
\begin{eqnarray} 
\Gamma_+(z)\alpha(X)=\alpha(X+\check z).
\label{multip1}
\end{eqnarray}

Let us repeat the preceding construction with another Pieri rule  defined in \cite{macdonald_symmetric_1999} p 215.
Let $\tilde \Omega(X)$ be the generating function of the elementary symmetric functions \cite{macdonald_symmetric_1999} (2.2) p 19:
\begin{eqnarray}
\tilde \Omega(X)=\Omega(-\epsilon X)=\prod_{x_k\in X}{(1+x_k)}
\label{Esymdef}
\end{eqnarray}
where the product $\epsilon X$ 
means we make the change of variable $x_i \to -x_i$ in $X$ and has a different meaning as $-X$ after \ref{Hsymdef}.
Following the conventions of \cite{boutillier_dimers_2015} 3.1, let us rename the previous operators $\Gamma_{L,\pm}$, and
similarly as in (\ref{multip}), define $\Gamma_{R,\pm}$ as:
\begin{eqnarray} 
\Gamma_{R,-}(z)\alpha(X)=\tilde \Omega(zX)\alpha(X),
\label{multiptilde}
\end{eqnarray}
and the same argument as above for $\Gamma_{R,+}(z)=\bar\Gamma_{R,-}(\check z)$ yields:
\begin{eqnarray}
\Gamma_{R,+}(z)\alpha(X)=\alpha(X-\epsilon \check z/(1-t)).
\label{multiptilde2}
\end{eqnarray}
It follows from these relations that $\Gamma_{a,+}(z)$ commutes with $\Gamma_{a',+}(w)$ and acts as the identity on the right vacuum, 
$\Gamma_{a,-}(z)$ commutes with $\Gamma_{a',-}(w)$ and acts as the identity on the left vacuum.
Moreover, the second line of (\ref{gammainfini}) generalizes to:
\begin{eqnarray} 
\Gamma_{a_1,+}(u)\Gamma_{a_2,-}(v)= \Gamma_{a_2,-}(v) \Gamma_{a_1,+}(u)
\left\lbrace
\begin{array}{lll}
  \Omega((1-t)v\check u) \\ 
 \tilde \Omega (v\check u)          \\
    \Omega((1-t)^{-1}v\check u)
    \end{array}
  \
    \begin{array}{lll}
     a_1=a_2=L \\  a_1\ne a_2 \\\ a_1=a_2=R
     \end{array}\right.
    \label{tildegammainfini}
\end{eqnarray}

Let $P^{},\ Q^{}$ denote the Hall-Littlewood Polynomials, and $P^{\omega},\ Q^{\omega}$ their $\omega$-duals
\cite{macdonald_symmetric_1999} chapter 7, section 5.1:
\begin{eqnarray}
P^{\omega}_{\lambda'}(V)&=&Q^{}_{\lambda}(-\epsilon V/(1-t))\nonumber\\
Q^{\omega}_{\lambda'}(V)&=&P^{}_{\lambda}(-\epsilon V/(1-t)).
\label{PQmacdo}
\end{eqnarray}
It is shown in particular that $P^{\omega}_{\lambda}$ is the  polynomial 
$P({\lambda,t,0})$ defined by (4.7) p 322,
as the symmetric functions decomposing triangularity:
$P^{\omega}_{\lambda}=m_\lambda+\sum_{\mu<\lambda}u_{\lambda,\mu} m_\mu$ on the symmetric  monomial basis,
and orthogonal for the scalar product dual to (\ref{scal101}) defined by: 
\begin{eqnarray} 
\langle \alpha(X)|\Omega(XY/(1-t))\rangle_\omega=\alpha(Y).
\label{scal101dual}
\end{eqnarray}

The states $|L,V\rangle$, $\langle L,\check U|$ are defined in (\ref{nouvX}), and
the states $|R,V\rangle$, $\langle R,\check U|$ by:
\begin{eqnarray} 
\langle R,\check U|&=&\tilde \Omega(\check UX)=\sum_\lambda P_{\lambda'}^{\omega}(\check U)\langle\bar\lambda|\nonumber\\
|R,{V}\rangle&=&\tilde \Omega(VX)=\sum_\lambda Q_{\lambda'}^{\omega}(V)|\lambda\rangle.
\label{nouvXdual}
\end{eqnarray}
Notice that restricting $\check U,V$ to $N$ variables has the effect to truncate the summation imposing
$\lambda_1\le N$, in other words all the sites greater than $N$ are empty.

Let $\lambda'_1\ge \lambda'_2\cdots \ge \lambda'_N\ge 0$ define  the partition conjugated to the q-boson partition $\lambda$:
$m_1=\lambda'_1-\lambda'_2,\cdots, m_{N-1}=\lambda'_{N-1}-\lambda'_N,m_{N}=\lambda'_{N}$.

 Let us introduce the following quantities attached to two partitions such that $\mu'\prec \lambda'$
(in other words, $\lambda-\mu$ is a vertical strip) \cite{macdonald_symmetric_1999} (3.2) p 215:
\begin{eqnarray}
\psi'_{\lambda/\mu}&=&\prod_{i\ge 1} {(\lambda'_i-\lambda'_{i+1})!_t\over (\lambda'_i-\mu'_{i})!_t(\mu'_i-\lambda'_{i+1})!_t}
\nonumber\\
\phi'_{\lambda/\mu}&=&\prod_{i\ge 1} {(\mu'_i-\mu'_{i+1})!_t\over (\lambda'_i-\mu'_{i})!_t(\mu'_i-\lambda'_{i+1})!_t}.
\label{psi'def}
\end{eqnarray}
Then, similarly as in (\ref{gauche1}), (\ref{gauche11}), the explicit expression of the transfer matrices $\Gamma_{R(z),\pm}$ acting 
on the semi infinite chains is:
\begin{eqnarray}
\Gamma_{R,+}(z)|\lambda\rangle
&=&\sum_{\mu'\prec \lambda'} \check z^{|\lambda-\mu|}\psi'_{\lambda/\mu}|\mu\rangle
\nonumber\\
\Gamma_{R,-}(z)|\mu\rangle
&=&\sum_{\mu'\prec \lambda'} z^{|\lambda-\mu|}\phi'_{\lambda/\mu}|\lambda\rangle.
\label{tildeGamma}
\end{eqnarray}
We shall derive these expressions  from (\ref{tildegammainfini}) using the Yang-Baxter equation in section (\ref{glomac1}).
Unlike $\Gamma_{L,\pm}$,
$\Gamma_{R,\pm}$ are not defined through a Lax matrix such as (\ref{laxq1}), but correspond to spin chains. 
If we attach  a decreasing sequence of nonnegative integers (spins)  $\mu'_1\ge\mu'_2\cdots\mu'_N$  to the  sites of a semi infinite chain,
the transfer matrix elements $(\Gamma_{R,-})_{\lambda\mu}$ are nonzero only when each spin $\mu'_k$ jumps in between his position 
and the position of the preceding spin: $\mu'_k\le\lambda'_k\le\mu'_{k-1}, \ (\mu'_0=\infty),$ so that
the number of nonzero spins can increase by one at each step. The matrix elements,
$\psi'_{\lambda/ \mu},\phi'_{\lambda/ \mu}$,
are the products of local Boltzmann weights between neighboring spins.

In particular, as in (\ref{droitt}) we have:
\begin{eqnarray} 
\Gamma_{R,-}(V)|\mu\rangle=\sum_{\lambda}Q^{\omega}_{\lambda'/\mu'}(V)|\lambda\rangle,
\label{tableaudual0}
\end{eqnarray}
from which, combining with (\ref{tildeGamma}),  we obtain the tableau expression for $Q^{\omega}_{\lambda'/\mu'}(V)$
analogous to (\ref{tableau}):
\begin{eqnarray} 
Q^{\omega}_{\lambda'/\mu'}(V)=\sum_T \phi'_T v^T
\label{tableaudual}
\end{eqnarray}
where this time:
\begin{eqnarray} 
\phi'_T=\phi'_{\lambda_0/\lambda_1}\cdots\phi'_{\lambda_{n-1}/\lambda_n}.
\label{droittdef}
\end{eqnarray}
For example, the one and two spin states are equal to:
\begin{eqnarray} 
|R,\{v_1\}\rangle&=&(S_1v_1)^{-1}_\infty\nonumber\\
|R,\{v_1,v_2\}\rangle&=&(S_1v_1)^{-1}_\infty(S_1v_2)^{-1}_\infty(S_2v_1v_2)^{-1}_\infty.
\label{tableaudual2spin}
\end{eqnarray}

Setting $\mu=0$ in (\ref{tableaudual0}), we obtain $| R,V\rangle $ which is an eigenstate of $\Gamma_{L,+}(z)$: 
\begin{eqnarray}  
\Gamma_{L,+}( z)| R,V\rangle=\tilde \Omega(V\check z)| R,V\rangle.
\label{gauche2-Toda}
\end{eqnarray}
As it will become clear in the next section, $\Gamma_{L,+}( z)$ is the transfer matrix of the open Toda chain,
and the Tableau expression of $| R,V\rangle $ coincides with the ``Baxter-Gauss-Givenal'' representation
of the q-Toda wave functions due to Gerasimov Lebedev and Oblezin \cite{gerasimov_q-deformed_2009}.

We can also recover the Product rule  \cite{macdonald_symmetric_1999}  p 215 by
applying $\Gamma_{R,+}(z)$ to $| L,V\rangle $: 
\begin{eqnarray}  
\Gamma_{R,+}( z)| L,V\rangle=\tilde \Omega(V\check z)| R,V\rangle,
\label{gauche2-Toda1}
\end{eqnarray}
and expanding in $z$:
\begin{eqnarray} 
\sum_{\lambda} \psi'_{\lambda/\mu}P_\lambda=e_r P_\mu,
\label{Hall}
\end{eqnarray}
where $\lambda-\mu$ is a vertical r-strip.

\section{Equivalence between the q-bosons and a Toda chain\label{todachain}}

In order to interpret the results of section (\ref{halfa}), we find it convenient to 
rephrase them in the  relativistic-Toda dynamics language \cite{ruijsenaars_relativistic_1990},
by demonstrating an equivalence between the two models.
We will be mainly concerned with a discretized version discussed in  
\cite{gerasimov_q-deformed_2008}\cite{gerasimov_q-deformed_2009}\cite{gerasimov_baxter_2013}
and recover the expression of their eigenvectors.

For this, we need to
to reinterpret the q-bosons configurations as Toda particle positions $\lambda'_i$
interacting with their nearest neighbors.
Define
\begin{eqnarray} 
X_k=S_{k}/S_{k-1}
\label{ydef}
\end{eqnarray}
($S_k=X_1\cdots X_{k}$),
and
for $1\le k\le N-1$:
\begin{eqnarray} 
\tau_{k}=x_{k}/x_{k+1}
\label{taudef}
\end{eqnarray}
obeying:
\begin{eqnarray} 
x_kX_k=tX_kx_k,
\label{xX=Xx}
\end{eqnarray}
and $x_k,X_l$ commute if the  labels $k$ and $l$ differ.

Setting:
\begin{eqnarray}
U_k&=&
\left( \begin{array}{cc}
1& x_k^{}\\S_{k}&.
\end{array}\right),
\label{ukdef0} 
\end{eqnarray}
and:
\begin{eqnarray}
L^{\rm Toda}_k&=&
\left( \begin{array}{cc}
1+zX_k&x_{k} \\-zX_kx_{k}^{-1}&.
\end{array}\right),
\label{ukdef} 
\end{eqnarray}
we have:
\begin{eqnarray} 
U_{k-1}L^{\rm Toda}_k=L_{k-1}U_k
\label{TodabosonL}
\end{eqnarray}
where $L$ is defined in (\ref{laxq1}).

Therefore, if we set:
 \begin{eqnarray} 
T^{\rm Toda}_N
=L_1^{\rm Toda}\cdots L^{\rm Toda}_{N},
\label{monodromytoda}
\end{eqnarray}
we have:
\begin{eqnarray} 
 U_0T^{\rm Toda}_N=T^{}_NU_N
\label{Tp10}
\end{eqnarray}
where $T_N$ is defined in (\ref{monodromy}).

\subsection{The Periodic chain}
The q-Toda chain transfer matrix \cite{ruijsenaars_relativistic_1990} is defined as:
\begin{eqnarray} 
 \Lambda_{N,n}^{\rm Toda}(z)={\rm tr}\ T^{\rm Toda}_{N}D^{\rm Toda}
\label{Tp100}
\end{eqnarray}
where:
\begin{eqnarray} 
D^{\rm Toda}=
\left( \begin{array}{cc}
1&  .\\.& t^n
\end{array}\right)
\label{periodcond}
\end{eqnarray}
is equal to the q-boson transfer Matrix $\Lambda_{N,x}(z)$ (\ref{perioddef}) restricted to the $n$-Boson sector if we require:
\begin{eqnarray} 
DU_0&=&U_ND^{\rm Toda}
\label{periodcond1}
\end{eqnarray}
with $D$ the q-boson twist $D=(1,X)$ (\ref{Ydef}).
Equivalently for:
\begin{eqnarray} 
x_{k-N}=x_{k}t^n&,&\ S_{k+N}=xS_k.
\label{periodcond2}
\end{eqnarray}
The periodic-Toda particles are labeled by $k, 1\le k\le N$ and their positions take discrete values:
$x_k=t^{\lambda'_k}$ with $\lambda'_k$ integer satisfying the periodicity conditions:
\begin{eqnarray} 
\lambda'_{k+N}=\lambda'_k-n.
\label{periodcond3}
\end{eqnarray}
The Boson occupation numbers are
equal to the difference between successive $\lambda$'s:
$m_k=\lambda'_k-\lambda'_{k+1}$, and
$X_k$ adds one unit to $\lambda'_k$: 
\begin{eqnarray} 
X_k|\lambda'_l\rangle=\delta_{k,l} |\lambda'_l+1\rangle.
\label{yk}
\end{eqnarray}
So, the periodic chain configuration space 
is characterized by $N$ integers $|\lambda'\rangle$ with  $ \lambda'_1\ge \cdots \ge \lambda'_N\ge  \lambda'_1-n$.

The momentum $x=X_1\cdots X_N$
translates the Toda positions $\lambda'_k$  globally by one unit:
\begin{eqnarray} 
|\lambda'_1 \cdots  \lambda'_N\rangle=x|\lambda'_1 -1\cdots  \lambda'_N-1\rangle,
\label{periodcond22}
\end{eqnarray}
In the sector where it is equal to  the q-boson twist parameter (\ref{Ydef}),
the periodic q-Toda chain transfer matrix (\ref{Tp100})  
coincides with the transfer matrix (\ref{TraceT}) restricted to the sector of $n$ Bosons.

In particular, the  $z$ coefficient is the right-mover Hamiltonian:
 \begin{eqnarray} 
H_1^{\rm per}&=&S_1\bar S_0+S_2\bar S_1+\cdots S_{N}\bar S_{N-1}\nonumber\\
&=&X_1(1-x_0/x_1)+X_2(1-x_1/x_2)+\cdots +X_N(1-x_{N-1}/x_N).
\label{Hamil}
\end{eqnarray}
As in (\ref{hermiteT}), we can redefine a Hermitian transfer matrix:
\begin{eqnarray} 
 \Lambda_{N,n}^{\rm Toda,H}(z)= \Lambda_{N,n}^{\rm Toda}(z)/\sqrt{xz^N}.
\label{Tp100hermite}
\end{eqnarray}
In particular, $x\bar H_1=H_{N-1}$, so $H_1$  commutes with his adjoint and both
can therefore be diagonalized simultaneously.

In the appendix (\ref{lieb}), using the equivalence with the q-boson problem, 
we obtain the Bethe equations for the spectrum and the corresponding eigenvectors which
are Hall-Littlewood polynomials specialized at the Bethe roots.
Notice however that the q-boson scalar product is associated to a  Hermitian conjugation different from the  one
considered by Ruijsenaars for $t$ a root of unity, namely $\bar X_k=X_k, \bar x_k=x_k$, and consequently, the diagonalization
problem we solve is different from his. The Q-Matrix approach described in
section (\ref{glomac0}) will be relevant in that case \cite{o._babelon_quantization_????}.

\subsection{The Open chain}
In the open case, we can as well represent $A_N$ and its hermitian conjugated $\bar A_N$ (\ref{ABinfini}) with the Toda variables.
We set the boundary variables to be: $S_0=1, x_0=0$, $S_{N+1}=\infty, x_{N+1}=1$, keeping the ratio $S_{N+1}/X_{N+1}=X_{1}\cdots X_{N}$ finite.
We consider the matrix element:
\begin{eqnarray} 
 A_N=(T^{B}_{N}U_{N+1})_{12}=(T^{\rm Toda}_{N})_{11},
\label{Antoda}
\end{eqnarray}
So, $A^{}_N$ is the open q-Toda chain  with $N$ particles conserved quantities generating function \cite{ruijsenaars_relativistic_1990}.
In particular, the coefficient of $z$ is the Hamiltonian:
 \begin{eqnarray} 
H_1^{\rm open}&=&S_1+S_2\bar S_1+\cdots S_{N}\bar S_{N-1}\nonumber\\
&=&X_1+X_2(1-x_1/x_2)+\cdots +X_N(1-x_{N-1}/x_N).
\label{Hamilouvert}
\end{eqnarray}
The Toda configurations $|\lambda'\rangle$, with  $\lambda'_1\ge \lambda'_2\cdots \ge \lambda'_N\ge 0$  
define  partitions conjugated to the q-boson partition,
$m_1=\lambda'_1-\lambda'_2,\cdots, m_{N-1}=\lambda'_{N-1}-\lambda'_N,m_{N}=\lambda'_{N}$. 

Similarly, 
$\bar A_N(z)=z^{-(N+1)}S_{N+1}^{-1}(T^{B}_{N}U_{N+1})_{11}$ (\ref{ABinfini}) is equal to:
\begin{eqnarray} 
 \bar A_N(z)=(\bar T^{\rm Toda}_{N}( z))_{11}-(\bar T^{\rm Toda}_{N}(z))_{12},
\label{Bntoda}
\end{eqnarray}
where we factorize $S_{N+1}/X_{N+1}=X_{1}\cdots X_{N}$ and {\it define}
$$\bar T^{\rm Toda}_{N}(z)=z^{-N}S_{N}^{-1}T^{\rm Toda}_{N}(z)$$
as in (\ref{monodromytoda}) with $\bar L_k^{\rm Toda}=(zX_k)^{-1}L_k^{\rm Toda}( z)$:
\begin{eqnarray} 
\bar L_k^{\rm Toda}(z)=
\left( \begin{array}{cc}
1+\check z\check X_k&\check z\check X_k x_k  \\-\check x_k& .
\end{array}\right)
\label{Lbardef}
\end{eqnarray}
Notice that $\bar T^{\rm Toda}_{N}(z) $ is {\it not} the hermitian conjugated of $T^{\rm Toda}_{N}(z)$.

Restricting $V$ to have $N$ nonzero variables, $V=V_N$, $\Gamma_{L,+}( z)$ 
acts in the space with the first $N$ sites occupied, and is equal to $\bar A_N( z)$.
From (\ref{gauche2-Toda}) we see that $| R,V_N\rangle$ are its eigenvectors with eigenvalue $\tilde \Omega(V_N\check z)$:
\begin{eqnarray}  
\bar A_N( z)| R,V_N\rangle=\tilde \Omega(V_N\check z)| R,V_N\rangle,
\label{gauche2-Toda2}
\end{eqnarray}
and thus we have obtained the left eigenvectors of the open Toda chain transfer matrix.

The open Toda spin chain   wave functions usually denote the (right) eigenvectors of $A_N(z)$ (\ref{gamma-infini}).
To properly define them, it is necessary to relax the constraint $\lambda'_N\ge 0$ 
and allow the spins
$\lambda'_k$ to take integer values in order
to be able to diagonalize $S_N$, its  $z^N$ coefficient. So, the definition of $\lambda'$ needs to be extended  
from partitions to sequences of decreasing integers.
The homogeneity property \cite{macdonald_symmetric_1999} (4.17) p 325:
\begin{eqnarray}  
P_{\lambda+1^N}(V_N)=v_1\cdots v_NP_{\lambda}(V_N)
\label{homog}
\end{eqnarray}
enables to extend the definition of $|R,V\rangle$. From (\ref{nouvXdual}):
$$\langle R,\check U|S_N=
\sum_\lambda P_{\lambda'+1^{N}}^{\omega}(\check U)\langle \bar\lambda|=\check u_1\cdots\check u_n\langle R,\check U|$$
we deduce:
\begin{eqnarray}  
\bar S_N|R,V\rangle=v_1\cdots v_N |R,V\rangle.
\label{homogextend}
\end{eqnarray}

We now can obtain the (right) eigenvectors of $A_N(z)$ (\ref{gamma-infini}) from those of $\bar A_N$ as follows. 
Let us denote $v=v_1\cdots v_N$ the eigenvalue of $\bar S_N$,
and $\bar A'_N(z,v)$ the restriction of $\bar A_N(z)$ to the space $S_N=v$. So, in $\bar A'_N(z,v)$ we substitute $v$ to $\bar S_N$. 
We can factor $|R,V\rangle=(vS_N)^{-1}_\infty|R',V\rangle$ to obtain the eigenstate $|R',V\rangle$ of
$\bar A'_N(z,v)$. 
Similarly, denoting $|W_h,V\rangle$ ($W_h$ for Whittaker)  the eigenstate of $A_N(z)$, and $A'_N(z,v)$ 
the restriction of $A_N$ to the space $A_N=v$ where we substitute $v$ to $ S_N$.
We can factor out $\delta(S_N/v)$
to obtain the eigenstate $|W_h',V\rangle$ of
$A'_N(z,v)$. 
We  recover $A'_N(z,v)$ when we substitute  $S_{N-k}/v$ to $S_{k}$  in the expression of  $\bar A'_N(\check z,v)$ .
From this, the eigenvector of the Toda chain with the eigenvalue $\tilde \Omega(\check U_N z)$
is given by:
\begin{eqnarray}  
|W_h,U\rangle=\delta(u_1\cdots u_NS_N)(u_1\cdots u_NS_N)_\infty|R,U\rangle,
\label{whittacker}
\end{eqnarray}
which means the inverse Pochhammer symbol must be replaced by a delta function.
For example, the two body wave function is obtained by replacing $(S_2v_1v_2)^{-1}_\infty$ by $\delta(S_2v_1v_2)$ in (\ref{tableaudual2spin}).

\section{Baxter Q Matrix \label{glomac}}
\subsection{Diagonalization of the closed Toda chain Transfer matrix\label{glomac0}}

In this section, we recover the periodic  q-boson, equivalently the closed discretized Toda-chain Bethe equations 
using Baxter Q-Matrix approach \cite{rodney_baxter_exactly_2008} chapter 9, or \cite{gaudin__michel_bethe_????}\cite{m._gaudin_fonction_1981}
chapter 8.6.
The results reported in this section have been obtained in collaboration with Olivier Babelon 
and Simon Ruisjenaars \cite{o._babelon_quantization_????}. 
The q-Toda Q-matrix (\ref{qdef}) was obtained in a previous work with Michel Gaudin \cite{m._gaudin_ansatz_1993}
and related Q-matrices appear in
recent works \cite{korff_cylindric_2012}\cite{zullo_q-difference_2015}\cite{korff_quantum_2015}.

The Bethe roots (\ref{bethef1}) and the corresponding eigenvalues (\ref{betheeigenvector10}) are recovered as
the solutions of the following divisibility problem (here specialized to the q-boson limit $s=0$):
\begin{eqnarray} 
\Lambda_N(z) q_n(z)=q_n(tz)+xz^Nt^nq_n(z/t),
\label{TQequation}
\end{eqnarray}
where $\Lambda_N(z)$ and $q_n(z)$ are respectively degree $N$ and $n$ polynomials. 
The strategy to diagonalize the closed chain transfer matrix  $\Lambda_{N,n}^{\rm Toda}(z)$ (\ref{Tp100})
consists in obtaining (\ref{TQequation}) as a matrix equation where $\Lambda_N(z)=\Lambda_{N,n}^{\rm Toda}(z),\ q_n(w)$ form a commuting family 
of polynomial matrices of the degree $N$ and $n$ respectively.
One then recovers (\ref{TQequation}) as the scalar equation  for their joined eigenvalues.

In simplest nontrivial case, $N=n=2$, the configuration space is three dimensional ($\lambda'_0=2,\ \lambda'_2=0 $) with
the basis $|\lambda'_1\rangle$ where $\lambda'_1$ takes the values $2,1,0$, and
the expressions of $\Lambda_2 (z)$ and $q_2(z)$ are given by:
\begin{eqnarray}
\Lambda_2 (z)=
\left( \begin{array}{ccc}
1+z^2x&(1-t)z & .\\(1-t^2)zx &1+z^2x &(1-t^2)z \\. &(1-t)zx &1+z^2x
\end{array}\right),
\label{exemplela} 
\end{eqnarray}

\begin{eqnarray}
q_2 (z)=
\left( \begin{array}{ccc}
1&-z& z^2\\-(1+t)zx & 1+z^2x &-(1+t)z \\z^2x^2&-zx &1
\end{array}\right).
\label{exempleq} 
\end{eqnarray}
Our aim here is to prove (\ref{TQequation}) in the general case and to relate the $q$-matrix to the Hall Pieri rules.

The first step is to obtain (\ref{TQequation}) as a vector equation. For this, we look for a null vector of a matrix conjugated to:
\begin{eqnarray} 
B_N=\left(T_D^{}\right)_{12},
\label{AnBn}
\end{eqnarray}
where $T_D^{\rm Toda}=T_N^{}D^{\rm Toda}$ and $T_N$ is defined as:
\begin{eqnarray} 
T^{}_N(z)
=L_1^{\rm Toda}\cdots L^{\rm Toda}_{2N+1},
\label{monodromytodaappendix}
\end{eqnarray}
and  $D^{\rm Toda}$
is the diagonal matrix: $(1,t^n)$ (\ref{periodcond}).
For future convenience, in the definition of $T_N$, we have labeled  
the Toda Lax matrices (\ref{ukdef}) with odd numbers.

We perform a similarity transformation on the $L_{2k+1}^{\rm Toda}$ matrices defining:
 \begin{eqnarray} 
L'_{2k+1}(z)=M_{2k}L_{2k+1}(z)M_{2k+2}^{-1},
\label{similarity}
\end{eqnarray}
where $M_{2k}$ is a triangular two by two matrix:
 \begin{eqnarray} 
M_{2k}=
\left( \begin{array}{cc}
1& x_{2k}\\.& 1
\end{array}\right),
\label{Mdef}
\end{eqnarray}
and $x_k=t^{\lambda'_k}$.
So, if we denote $T'_D$ the matrix obtained by substituting $L'_{2k+1}$ to $L_{2k+1}$ in $T_D$, 
we have $T'_D=M_0T_DM_0^{-1}$.

We look for a null eigenvector $\psi_{2k+1}$ of $(L'_{2k+1})_{12}$, which amounts to the condition:
 \begin{eqnarray} 
(zX_{2k+1}(1-{x_{2k}\over x_{2k+1}})+(1-{x_{2k+1}\over x_{2k+2}}))\psi_{2k+1}=0,
\label{psiequ}
\end{eqnarray}
with the solution:
 \begin{eqnarray} 
\psi_{2k+1}=\sum_{\lambda'_{2k+1}}\psi_{\lambda'_{2k+1}}^{\lambda'_{2k},\lambda'_{2k+2}}(-z)|\lambda'_{2k+1}\rangle,
\label{psisol}
\end{eqnarray}
where the sum is for $\lambda'_{2k}\ge\lambda'_{2k+1}\ge\lambda'_{2k+2}$,
and:
\begin{eqnarray}
\psi_{b}^{a,c}(z)={z}^{b-c}
{a-c\choose a-b}_t
\label{psisol1}
\end{eqnarray}
with $a\ge b\ge c$. 
Then, evaluating the action of the diagonal elements yields:
 \begin{eqnarray} 
(L'_{2k+1})_{11}\psi_{2k+1}(z)&=&\psi_{2k+1}(tz)
\nonumber\\
(L'_{2k+1})_{22}\psi_{2k+1}(z)&=&zX_k\psi_{2k+1}(z/t).
\label{diago}
\end{eqnarray}

So,
given a set of decreasing integers $\lambda'_{0}\ge\lambda'_{2}\ge\cdots\ge \lambda'_{2N-2}\ge \lambda'_{0}-n$, 
imposing the periodicity condition 
\begin{eqnarray} 
M_0=M_{2N}D^{\rm Toda},
\label{conditionperiodic}
\end{eqnarray}
equivalently $\lambda'_{2N}=\lambda'_{0}-n$ 
a solution of:
\begin{eqnarray} 
(T'_D)_{12}\psi^{\{\lambda'_{2k}\}}=0
\label{psisol10}
\end{eqnarray}
is given by:
\begin{eqnarray} 
\psi^{\{\lambda'_{2k}\}}=\psi_1\otimes\psi_3\cdots\otimes\psi_{2N-1}.
\label{psisol2}
\end{eqnarray}
Due to the triangularity of $L'_{2k+1}\psi_{2k+1}$:
\begin{eqnarray} 
L'_{2k+1}\psi_{2k+1}=
\left( \begin{array}{cc}
\psi_{2k+1}(tz)&  .\\ \star & zX_{2k+1}\psi_{2k+1}(z/t)
\end{array}\right)
\label{tildeLaction}
\end{eqnarray}
the action of $T'_D$ on $\psi^{\{\lambda'_{2k}\}}$ is a lower triangular two by two matrix
the diagonal elements of of which are the  respectively the direct product of the Lax matrix diagonal elements.
Taking its trace, 
we recover (\ref{TQequation}) as a set of vector equations when we substitute $\psi^{\{\lambda'_{2k}\}}$ to $q$.

The second step is to organize the vectors $\psi^{\{\lambda'_{2k}\}}$ into a matrix, by requiring the commutation:
\begin{eqnarray} 
\Lambda(z) q(z)=q(z)\Lambda(z) 
\label{TQequationcommut}
\end{eqnarray}
For this, let us rewrite (\ref{TQequation}) in a manifestly conjugate invariant way:
\begin{eqnarray} 
\Lambda^{\rm H} q^{\rm H}(z)=(t^n/xz^N)^{1/2}q^{\rm H}(tz)+(t^nxz^N)^{1/2}q^{\rm H}(z/t)
\label{TQequation1}
\end{eqnarray}
where $q=z^{n/2}q^{\rm H}$.
So, due to the fact that two hermitian matrices commute if and only if
their product is hermitian, (\ref{TQequationcommut}) will hold if we can define $q^{\rm H}$ hermitian obeying (\ref{TQequation1}).

In the Toda notations, the scalar product (\ref{scal1}) is given by:
\begin{eqnarray} 
\langle \mu'_1\cdots  \mu'_N|\lambda'_1\cdots  \lambda'_N\rangle&=& \delta_{\lambda',\mu'} \prod_1^{N} ({\lambda'_k-\lambda'_{k+1})!_t}.
\label{scal1todaperiod}
\end{eqnarray}
For example, the norm of the states $\lambda'_1=2,1,0$ in the $N=n=2$ case are respectively $(t)_2,(t)_1^2,(t)_2 $.

Consider the matrix $q$ defined by:
\begin{eqnarray} 
&&q(z)=\sum_{\{\lambda'_{2l}\}} \psi^{\{\lambda'_{2l}\}}\langle \lambda'_{2l}|=\nonumber\\
&&\sum_{\{\lambda'_{l}\}}| \lambda'_{2l+1}\rangle 
\prod_{k=0}^{N-1}{(-z)^{\lambda'_{2k}-\lambda'_{2k+1}}\over(\lambda'_{2k}-\lambda'_{2k+1})!_t (\lambda'_{2k+1}-\lambda'_{2k+2})!_t }
\langle \lambda'_{2l}|
\label{qdef}
\end{eqnarray}
where the sum is on decreasing sequences of integers $\lambda'_{0}=n\ge \lambda'_{1}\ge \cdots \ge \lambda'_{2N}=0\ge \lambda'_{2N+1}=\lambda'_{1}-n$.
We can bring $\lambda'_1$ to be equal to $n$ in the bra using (\ref{periodcond22}):
\begin{eqnarray} 
|\lambda'_1 \cdots  \lambda'_{2N-1}\rangle=x^{-\lambda'_{2N+1}}|n,\lambda'_3 -\lambda'_{2N+1},\cdots , \lambda'_{2N-1}-\lambda'_{2N+1}\rangle.
\label{periodcond22b}
\end{eqnarray}
It defines a degree $n$ polynomial matrix in $z$, the zero degree matrix elements being for $\lambda'_{2k+1}=\lambda'_{2k}$, and
the maximal degree elements for  $\lambda'_{2k}=\lambda'_{2k-1}$. 
It commutes with the translation $\mathcal{{T}}$ (\ref{translate})
acting on Toda configurations as:
\begin{eqnarray} 
\mathcal{{T}}|\lambda'_k\rangle=|\lambda'_{k+2}\rangle
\label{Cpermut}
\end{eqnarray}
\begin{eqnarray} 
q^{}\mathcal{{T}}=\mathcal{{T}}q^{}
\label{TQequationhermite}
\end{eqnarray}
and the matrix $q^{\rm H}=z^{-N/2}q$ 
obeys the hermitian relation:
\begin{eqnarray} 
\mathcal{T}\bar q^{\rm H}=q^{\rm H}
\label{TQequationhermite2}
\end{eqnarray}
Since $\mathcal{{T}}$ commutes with $\Lambda$, (\ref{TQequationcommut}) is satisfied.
Moreover, since all matrices commute with the momentum $x$, the commutation holds
in each finite dimensional momentum sector.

Returning to standard notations: $\mu'_k=\lambda'_{2k+1}, \nu'_{k+1}=\lambda'_{2k}$,
$q(z)$ is  given by an expression similar to (\ref{tildeGamma}):
\begin{eqnarray}
q(z)|\nu\rangle
&=&\sum_{\mu} (-z)^{\sum\nu'_k-\mu'_k}\prod_{k=1}^N{\nu'_k-\nu'_{k+1}\choose \nu'_k-\mu'_{k+1}}_t|\mu\rangle
\label{tildeGammaqq}
\end{eqnarray}
where the sequences $\nu,\mu$ are interlaced as: $\nu'_1=n\ge\mu'_1\cdots\ge\mu'_N\ge 0$.

Last, we must show that $q(u)$  commutes  with $\Lambda(z)$ and $q(v)$. 
Here, we only show the first commutation by adapting 
an argument of Sklyanin  \cite{sklyanin_backlund_2000}.

Denote $\mathcal{N}$ the diagonal matrix of the norms:
\begin{eqnarray} 
\mathcal{N}|{\mu}\rangle=\langle \mu|\mu\rangle|{\mu}\rangle.
\label{Nnorm}
\end{eqnarray}
Given a matrix $\Lambda$, we denote its conjugated by  $\mathcal{N}$ as: 
\begin{eqnarray} 
\tilde \Lambda=\mathcal{N}\Lambda\mathcal{N}^{-1},
\label{Nconjug}
\end{eqnarray}
and so,
the commutation relation to be shown becomes:
\begin{eqnarray} 
\Lambda(z)q(u)\mathcal{N}^{-1}=q(u)\mathcal{N}^{-1}\tilde\Lambda( z)
\label{commutskly}
\end{eqnarray}

To show this relation, let us introduce a Lax operator
$\mathbb{L}(z)$ and the corresponding monodromy matrix:
\begin{eqnarray}
\mathbb{T}_N(z)=\mathbb{L}_1(z)\cdots \mathbb{L_N}(z)
\label{monodromyQ} 
\end{eqnarray}
to represent $q_N(z)$ as:
\begin{eqnarray} 
q_N(z)\mathcal{N}^{-1}
={\rm tr}_s(\mathbb{T}_N(-z)D^{s}),
\label{monodromytodaappendixQ}
\end{eqnarray}
and use (\ref{Tp100}) to represent $\Lambda$. 

The condition:
\begin{eqnarray}
R_{\sigma s}D^{\sigma}D^s=D^{\sigma}D^sR_{\sigma s}
\label{qsolu2} 
\end{eqnarray}
where $D^{\sigma}={\rm diag}(1,t^n)$ determines $D^s$ (\ref{monodromytodaappendixQ}) to be:
\begin{eqnarray}
D^s=S^{-n}
\label{qsolu22} 
\end{eqnarray}

From the definition of the hermitian conjugation, $\tilde \Lambda={^t{\bar \Lambda}}(\check z)$, 
is obtained by substituting $\tilde L^{\rm Toda}(z)$ to $L^{\rm Toda}(z)$ in (\ref{monodromytodaappendix}),
where $\tilde L^{\rm Toda}_k(z)= {^t\bar L^{\rm Toda}_k}(\check z) $ with $\bar L^{\rm Toda}$ defined in (\ref{Lbardef}):
\begin{eqnarray}
\tilde L^{\rm Toda}_k(z)&=&
\left( \begin{array}{cc}
1+zX_k&zx_{k}X_k \\-x_{k}^{-1}&.
\end{array}\right).
\label{l'todadef} 
\end{eqnarray}

The commutation relation will then follow from the intertwining Yang-Baxter relation \cite{faddeev_how_1996}:
\begin{eqnarray}
R_{\sigma s}(z/u)L^{\rm Toda}_{\sigma x}(z)\mathbb{L}_{s x}(u)=\mathbb{L}_{s x}(u)\tilde L^{\rm Toda}_{\sigma x}(z)R_{\sigma s}(z/u)
\label{YBq} 
\end{eqnarray}
The notation $L_{\sigma x},\ \mathbb{L}_{s x},$ 
means the matrices act respectively  in the direct product of the two dimensional $\sigma$ space 
and the $x$ spin space, the direct product of the $s$ and $x$ spin spaces.

Multiplying (\ref{YBq}) by the permutation $P_{x s}$ of $x$ and $s$ to the right, it rewrites:
\begin{eqnarray}
R_{\sigma s}(z/u)L^{\rm Toda}_{\sigma x}(z) {\check {\mathbb L}}_{}(u)=\mathbb{\check L}_{}(u)\tilde L^{\rm Toda}_{\sigma s}(z)R_{\sigma x}(z/u)
\label{YBq2} 
\end{eqnarray}
where $ \mathbb{\check L}= \mathbb{ L}_{sx}P_{sx}$.
If we take:
\begin{eqnarray}
R_{\sigma s}(z)=
\left( \begin{array}{cc}
1+zS&s \\-s^{-1}(1-S)&-1
\end{array}\right),
\label{Rtodadef} 
\end{eqnarray}
This R-matrix is the q-deformed R-matrix of \cite{gaudin__michel_bethe_????} (14.54) used by Sklyanin \cite{sklyanin_backlund_2000}
in the Toda case,
it also appears in Korff in a similar context \cite{korff_cylindric_2012}.
The above relation rewrites in matrix form as:
\begin{eqnarray}
&&\left( \begin{array}{cc}
1+z/uS&s \\-s^{-1}(1-S)&-1
\end{array}\right)
\left( \begin{array}{cc}
1+zX&x \\-zXx^{-1}&.
\end{array}\right)
\mathbb{\check L}_{}(u)=
\nonumber\\
&&
\mathbb{\check L}_{}(u)
\left( \begin{array}{cc}
1+zS&zsS \\-s^{-1}&.
\end{array}\right)
\left( \begin{array}{cc}
1+z/uX&x \\-x^{-1}(1-X)&-1
\end{array}\right)
\label{qRL} 
\end{eqnarray}
which amounts to the following relations:
\begin{eqnarray}
\mathbb{\check L}x&=&x\mathbb{\check L}
\nonumber\\
\mathbb{\check L}SX&=&SX\mathbb{\check L}
\nonumber\\
\mathbb{\check L} {x/ s}&=&{x/s}(1-S)\mathbb{\check L}
\nonumber\\
u\mathbb{\check L}(1-{s/ x})S&=&S\mathbb{\check L}
\label{Rtodacommut} 
\end{eqnarray}
to be satisfied with: $sS=tSs$ and $xX=tXx$.
The solution is factorized as:
\begin{eqnarray}
\mathbb{\check L}_{}(u)=G(S)F_u(s/x)
\label{qsolu} 
\end{eqnarray}
where $F,G$ satisfy Pochammer (q-Gamma) identity:
\begin{eqnarray}
uF_u(s)(1-s)S&=&SF(s)\nonumber\\
(1-S)G(S)s&=&sG(S)
\label{qsolu1} 
\end{eqnarray}
and are represented as follow in the basis where $s$ is diagonal $s|n\rangle=t^n|n\rangle$:
\begin{eqnarray}
F_u|n\rangle&=&\check u^n/n!_t|n\rangle\nonumber\\
G|n\rangle&=&\sum_{k\ge n}1/(k-n)!_t|k\rangle
\label{qsolu21} 
\end{eqnarray}
In matrix form:
\begin{eqnarray}
\mathbb{ L}_{sx}(u)|\nu_s,\nu_x\rangle=\sum_{\mu_s}{\check u^{\nu_x-\nu_s}\over (\mu_s-\nu_x)!_t(\nu_x-\nu_s)!_t}|\mu_s,\nu_s\rangle,
\label{qsolumatrix} 
\end{eqnarray}
with $\mu_s\ge \nu_x\ge \nu_s$.

Substituting (\ref{qsolumatrix}), (\ref{qsolu22}) in (\ref{monodromytodaappendixQ}), 
we verify that $q(z)$  agrees with our previous expression (\ref{tildeGammaqq}),
concluding the proof that $q(z)$ and $\Lambda(w)$ commute.

\subsection{The  Hall  Pieri rule\label{glomac1}}

In this section, we obtain the expression of the vertex operators $\Gamma_{R,\pm}$ (\ref{tildeGamma})
by solving the two hermitian  conjugated relations:
\begin{eqnarray} 
\Gamma_{R,+}(u)\Gamma_{L,-}(z)&=& (1+z\check u)\Gamma_{L,-}(z) \Gamma_{R,+}(u)\nonumber\\
\Gamma_{L,+}( z)\Gamma_{R,-}( u)&=& (1+\check z u)\Gamma_{R,-}( u) \Gamma_{L,+}( z)
\label{tildegammainfiniB}
\end{eqnarray}
The first equation tells us that if we
iterate the action of $\Gamma_{L,-}(v_i)$ on the vacuum, we create the state $|L,V\rangle$ eigenstate of $\Gamma_{R,+}(z)$ with the
eigenvalue  $\tilde\Omega(\check z V)$. It
is therefore equivalent to the Hall-Pieri rule (\ref{Hall}). 
The second equation tells us that if we iterate the action of $\Gamma_{R,-}(v_i)$ on the vacuum, 
we create the $\omega$-dual states $|R,V\rangle$, eigenstates of $\Gamma_{L,+}( z)$ with the same eigenvalue $\tilde\Omega(\check z V)$,
closely connected to the open Toda chain eigenstates as we saw in section (\ref{todachain}).

Let $P_N$ denote the projector onto partitions $|\lambda\rangle$ having the multiplicity
of $1$, $\lambda_1,$  less or equal to $N$: $\lambda'_{N+1}=0$, and $P^M$ the projector onto partitions of length $l(\lambda)=\lambda'_1$ less or equal to $M$.
By construction, $\Gamma_{L,-}P^M=P^{M+1}\Gamma_{L,-}P^M$, and the finite size q-boson transfer matrices defined in   (\ref{ABinfini}) are
obtained by projection: $A_N^L=P_N\Gamma_-^L=P_N\Gamma_-^LP_N$. 

We shall require $\Gamma_{R,\pm}$ to be  hermitian conjugated:
\begin{eqnarray} 
\bar\Gamma_{R,-}(z)=\Gamma_{R,+}( z).
\label{Gamma-conjug1}
\end{eqnarray}
Define the finite size projections of $\Gamma_{R,+}$ to be $\bar A_N^R$
where $\bar A_N^R=P_N\Gamma^R_+=P_N\Gamma^R_+P_{N+1}$,
and require $\Gamma^R_+P_M=P_M\Gamma^R_+P_M$.

With these condition satisfied, the two equations (\ref{tildegammainfiniB}) are conjugated.
Dividing the first one by the norm matrix ($\mathcal N$) to its right and projecting with $P_{N}$ to its left, it becomes:
\begin{eqnarray} 
 (1+z\check u) A_N^L(z)\bar A^R_N(u){\mathcal N}^{-1}=\bar A^R_N(u){\mathcal N}^{-1}\tilde A_{N+1}^L(z)
\label{tildegammaproject1}
\end{eqnarray}
where:
\begin{eqnarray} 
A_{N}^L(z)&=&T^{\rm Toda}_{N}( z)_{11}\nonumber\\
\tilde A_{N+1}^L(z)&=&\tilde T^{\rm Toda}_{N+1}( z)_{11}-\tilde T^{\rm Toda}_{N+1}(z)_{12}.
\label{Nconjug1}
\end{eqnarray}

We take $\bar A^R_N(u)$ equal to:
\begin{eqnarray} 
\bar A^R_N(u)= (\mathbb{T}_{N+1}(u))_{\infty 0}{\mathcal N}.
\label{BRudef}
\end{eqnarray}
In matrix form:
\begin{eqnarray}
\bar A^R_N(u)|\lambda
\rangle=\sum_{\mu}\prod_{k=1}^{N+1}\check u^{\lambda_k-\mu_k}{ (\lambda_k-\lambda_{k+1})!_t\over (\lambda_k-\mu_k)!_t(\mu_k-\lambda_{k+1})!_t}|\mu\rangle,
\label{Bqsolumatrix} 
\end{eqnarray}
with $\lambda_1\ge \mu_1\cdots \ge \mu_N\ge\lambda_{N+1}\ge\mu_{N+1}=0$.
$\bar A^R_N(u)$ satisfies the required commutation relations with $P^M$ and $P_N$, 
and the corresponding $N=\infty$ matrix $\Gamma_{R,+}$ agrees with the definition (\ref{tildeGamma}) of the 
Hall Pieri rules.

To prove (\ref{tildegammaproject1}), consider the Matrix element of the Yang-Baxter equation
(\ref{YBq}):
\begin{eqnarray} 
R_{\sigma s}(z\check u)(T_{N+1}^{\rm Toda})_{\sigma }(z)(\mathbb{T}_{N+1})_{s}(u)=
(\mathbb{T}_{N+1})_{s }(u)(\tilde T_{N+1}^{\rm Toda})_{\sigma }(z)R_{\sigma s}(z\check u)
\label{YBTT}
\end{eqnarray}
taken between the states $\langle 1|\otimes \langle \infty|$ and $| 2\rangle\otimes | 0\rangle$ 
in the direct product of the two dimensional $\sigma$-space with the spin $s$-space.
We eliminate $R$ on the left-hand side with  $\langle 1|\otimes \langle \infty|R=(1+z\check u)
\langle 1|\otimes \langle \infty|$, and on the-right hand side with
$R| 2\rangle\otimes | 0\rangle=(| 1\rangle-|  2\rangle)\otimes | 0\rangle$. We also remove $L^{\rm Toda}_{N+1}$
on the left-hand side using
$L_{N+1}^{\rm Toda}|2\rangle=|1\rangle$, due to the zero value of the boundary spin: $\mu_{N+1}=0$.
The resulting expression is equal to (\ref{tildegammaproject1}).

Notice that unlike the  commutation relations between $\Gamma_{L,-}(z)$ and $\Gamma_{L,+}(u)$, 
the  commutation relations between $\Gamma_{L,-}(z)$ and $\Gamma_{R,+}(u)$
are already satisfied in finite size. The commutations between $\Gamma_{R,\pm}$ and   $\Gamma_{L,\pm}$
can be obtained along the same line and we shall study the commutations of $\Gamma_{R,\pm}$ between themselves elsewhere
\cite{antoine_duval_phd_????}.

\section{A remark on the Half infinite chain and the Gaudin determinant\label{gauddet}}
In this section, we consider the scalar product of two XXZ Bethe states on a semi-infinite
chain
which Gaudin expressed using the determinant  (\ref{Delta-det}) \cite{gaudin_michel_bose_1971}.
Recently, Kirillov and Noumi \cite{kirillov_affine_????}, and Warnaar \cite{warnaar_bisymmetric_2008} 
obtained the same determinant by applying the Sekuiguchi-Debiard operator to
the Cauchy identity (\ref{cauchy}). Betea and Wheeler \cite{betea_refined_2016} reinterpreted it as the scalar product of
two q-boson wave functions on the semi-infinite
chain. We present here an interpolating scalar product (We do not give the proof here, but the result can already be inferred from the two particle case)
by extending the argument of \cite{m._gaudin_fonction_1981}\cite{gaudin__michel_bethe_????} chapter 4. Appendix B 
to an arbitrary spin. 

We now consider the $SL_2$ lax matrix reducing to (\ref{laxq1}) for $s=0$:
 \begin{eqnarray}
(1+zs)L^s(z)=
\left( \begin{array}{cc}
1+zK&  z  S^-\\S^+& z+K
\end{array}\right)
\label{lax-sl2} 
\end{eqnarray}
where:
\begin{eqnarray}
K=s\tau,\ S^-=S^{-1}(1-\tau),\ S^+=S(1-s^2\tau).
\label{spin-sl2} 
\end{eqnarray}

The scalar product such that $S^-$ is the adjoint of $S^+$:
\begin{eqnarray} 
\langle {n}|{m}\rangle_s&=& (t)_m/(s^2)_m\delta_{n,m}
\label{scalS}
\end{eqnarray}
replaces (\ref{scal}).

The Bethe-Lieb eigenvectors of the semi-infinite chain
take the form (\ref{betheeigenvector}):
\begin{eqnarray}
R^s_\mu(\check U_n)={1\over \prod_{k=1}^n(1+s\check u_k)}
\sum_P B(\check u_i) \xi(\check u_1)^{\mu_1}\cdots   \xi(\check u_n)^{\mu_n}
\end{eqnarray}
where $P$ permutes the $\check u_k$, and  $ \xi(\check u_k)$, $B(\check u_i)$, are defined in (\ref{betheeigenvector5}), (\ref{betheeigenvector7}).
The normalization prefactor is put for further convenience.

Notice that $R^s_\mu$ is proportional to the symmetrizer of the Hecke algebra $\Cup$ acting on the monomial $\prod_1^n \xi(\check u_i)^{\mu_i}$
where  $\Cup$ defined as:
\begin{eqnarray} 
f\Cup(\check u_l)={(1-t)^n\over n!_t}\sum_P f(\check u_1,\cdots,\check u_n)\prod_{i<j} {\check u_i-t\check u_j\over \check u_i-\check u_j}
\label{varpi}
\end{eqnarray}
and the Hecke algebra is generated by:
\begin{eqnarray} 
f\Cup_k(\check u_l)=f(\cdots \check u_k,\check u_{k+1}\cdots){\check u_k-t\check u_{k+1}\over \check u_k-\check u_{k+1}}
+(\check u_k\leftrightarrow \check u_{k+1})
\label{varpi1}
\end{eqnarray}

We consider the scalar product of two Bethe wave functions on the semi infinite line $\mu_k\ge 0$:
\begin{eqnarray}
\Delta^s_n(\check U_n,V_n)=\sum_{0\le \mu_1\le\cdots\le \mu_n}{1\over \langle \mu|\mu\rangle_s}R^s_\mu(\check U_n)R^s_{\mu}(V_n).
\label{scqlarproductS}
\end{eqnarray}
It is given by:
\begin{eqnarray}
\Delta_n^s(\check U_n,V_n)={t^{n(n-1)\over 2}\over (1-t)^n}{D_n\over \delta_n} 
\label{Delta-scal}
\end{eqnarray}
where $\delta_n$ is the Cauchy determinant:
\begin{eqnarray}
\delta_n=|{1\over 1-t{\check u_k v_l}}|,
\label{delta-det}
\end{eqnarray}
$\Delta_n$  the Gaudin determinant:
\begin{eqnarray}
D_n=|{1\over(1-{\check u_k v_l}) (1-t{\check u_k v_l})}|.
\label{Delta-det}
\end{eqnarray}
Therefore, the scalar product which interpolates between the XXZ-chain for $ts^2=1$ 
and the Hall-Littlewood polynomials  for $s=0$ does not depend on  the spin.

Kirillov, Noumi and Warnaar obtained the q-deformation of the Hall-Littlewood formula as:
\begin{eqnarray}
\sum_{\lambda} P^{qt}_\lambda(V_n)Q^{qt}_\lambda(\check U_n)
\prod_{k=1}^n(1-q^{\lambda_k}t^{n-k+1})=\Omega({1-t\over 1-q}qV_n\check U_n){t^{n(n-1)\over 2} (1-t)^n}
{D_n\over \delta_n}
\nonumber\\
\label{warnaar}
\end{eqnarray}
where $P^{qt}$ is the Macdonald polynomial \cite{macdonald_symmetric_1999} and $\lambda_1\ge\cdots \ge\lambda_n\ge 0$.
Indeed, if one sets $q=0$ in the preceding expression, using (\ref{rtoq}), it rewrites:
\begin{eqnarray}
\sum_{\lambda} {R_\lambda(V_n)R_\lambda(\check U_n)\over \langle\lambda|\lambda\rangle}={t^{n(n-1)\over 2}\over (1-t)^n}{D_n\over \delta_n}
\label{warnaarlascoux}
\end{eqnarray}
which is (\ref{Delta-scal}) for $s=0$. 

Lascoux \cite{lascoux_gaudin_2007} reduced the $q\ne 0$ case to $q=0$ as follows.
The only $q$ dependence of the right hand side of (\ref{warnaar}) comes from the first factor, and we can also factor it out from the left hand side.
Representing the product
as the eigenvalue of a Dunkl operator \cite{bernard_yang-baxter_1993}) enables to rewrite the left hand side as a modification of the Cauchy Kernel equal to:
\begin{eqnarray}
\Omega({1-t\over 1-q} \check U_nV_n)(1-t\tau_q)\cdots(1-t^n\tau_q)\Cup
\label{lascoux}
\end{eqnarray}
with $\check u_i\tau_q=\check u_{i+1}$ and $\check u_{n+1}=q\check u_1$.

Let $X_n=X_k+\bar X_k$ where 
$X_k=\{x_1,\cdots x_k\}$, $\bar X_k=\{x_{k+1},\cdots x_n\}$. We have:
$$\Omega({1-t\over 1-q} X_n)\tau_q^k=\Omega({1-t\over 1-q} (qX_k+\bar X_k))$$
where $\tau_q$ is defined as above with $\check u \to x$.
Substituting $qX_k+\bar X_k=qX_n+(1-q)\bar X_k$, we deduce it is also equal to:
\begin{eqnarray}
\Omega(({1-t}) X_n)\tau_0^k\Omega(q{1-t\over 1-q}X_n).
\label{warnaarlascoux2}
\end{eqnarray}
So, (\ref{lascoux}) is also equal to:
\begin{eqnarray}
\Omega({(1-t)}V_n\check U_n)(1-t\tau_0)\cdots(1-t^n\tau_0)\Cup \Omega({1-t\over 1-q}qV_n\check U_n)
\label{lascoux2}
\end{eqnarray}
reducing $q\ne 0$ to:
\begin{eqnarray}
\Omega({(1-t)}V_n\check U_n)(1-t\tau_0)\cdots(1-t^n\tau_0)\Cup=t^{n(n-1)\over 2}(1-t)^n{D_n\over \delta_n}
\label{lascoux3}
\end{eqnarray}
which is  (\ref{warnaarlascoux}).

\section{Conclusions and perspectives}

In this paper, we have studied the q-boson model and a discretized version of the Toda lattice on finite and semi infinite chains
in relation with the theory of the Hall-Littlewood multi-variable orthogonal polynomials. We have observed
that the semi infinite transfer matrices are vertex operators with Cauchy type commutation relations.

As a very interesting application, we observe that the q-boson model appears via the Hall-Pieri rules
when studying the current fluctuations of the Asymmetric Simple Exclusion Process, as explained by
Lazarescu in part III of his thesis \cite{lazarescu_exact_2013}.

Although most of the semi infinite chain results are straightforward to extend to the Macdonald case, the underlying statistical models
seem not to be expressible in terms of local Boltzmann weights. However, some important simplifications occur
with the Hall-Pieri rules \cite{garsia_new_2014} which suggest that these Macdonald lattice models are simpler than they appear.
One can also run
the process backwards to define them on finite lattices. It will be interesting to investigate
how the finite size version is connected to the Yang-Baxter equation and 
the possibility to use the Bethe-ansatz to obtain its spectrum.

Another point not touched here is the fact that the components of the eigenvectors
are also eigenstates of the Sekuiguchi-Debiard operators.

The very simple discretized Toda chain we have introduced here can be diagonalized by coordinate Bethe-Anzats. It
does not satisfy the more physical Hermitian condition of the Ruijsenaars q-Toda chain. Although we cannot use
the coordinate Bethe-ansatz in that case, we expect that the Q-matrix
diagonalization of section (\ref{glomac}) reveals useful as for the Toda chain \cite{pasquier_periodic_1992}
(see for example \cite{kharchev_unitary_2002}\cite{o._babelon_quantization_????}).

\section*{Acknowledgement}

V. Pasquier is grateful to Junichi Shiraishi for numerous friendly discussions, in particular for explaining him 
the Pieri rules.

We also thank  Olivier Babelon, Jérémie Bouttier, Sylvie Corteel, Alexandre Lazarescu and Simon Ruisjenaars 
for illuminating discussions.

Part of this work was done during the workshop  "Beyond Integrability" held at the CRM,  Montreal.


\begin{appendix}

\section{Bethe-Lieb diagonalization\label{lieb}}
In this appendix, we partially diagonalize the transfer matrix (\ref{perioddef}) of the spin chain considered in section (\ref{gauddet})
by adapting the discussion of \cite{m._gaudin_fonction_1981} \cite{gaudin__michel_bethe_????} 
section 7.3. or \cite{rodney_baxter_exactly_2008} sections 8.2-4 to this case.
The Bethe-Lieb technique has the advantage to provide an
explicit expression of the eigenfunctions  (which are studied thoroughly in \cite{borodin_family_2014} in the infinite chain limit).

 The monodromy matrix is:
 \begin{eqnarray} 
T^s_N(z)
=L^s_0L^s_1\cdots L^s_{N-1}
\label{monodromys}
\end{eqnarray}
where $L^s$ is defined by (\ref{lax-sl2}), 
we want to find the left eigenvectors of:
 \begin{eqnarray} 
\Lambda^s_{N,X}(z)={\rm  Tr}\ T^s_N(z)D
\label{perioddefs}
\end{eqnarray}
with $D=(1,X)$  (\ref{Ydef}).

In the sector of $M$ Bosons, the eigenvalue equation writes:
\begin{eqnarray}
\sum_\lambda R_\lambda(  U)(\Lambda^s_N)_{\lambda\mu}(z)=\tilde\Lambda_N^s (z U)R_\mu(  U)
\label{betheeigen2s}
\end{eqnarray}
where $l(\mu)=l(\lambda)=M$.
We limit our treatment to the states $\mu$ such that the lattice sites are  occupied
by at most one Boson. All the parts $\mu_k$ are distinct: $m_i({\mu})\le 1$. Therefore,
the only $\lambda$ which contribute to the sum are those for which $m_i({\lambda})\le 2$.

We define the Bolzman weights:
\begin{eqnarray}  
\begin{array}{llll}
\omega_1=1+zs&\omega_3={z+s}&\omega_5={z(1-t)}\\
\omega_2={z+ts}&\omega_4={1+zts}&\omega_6={1-s^2}&\omega_0=1-s^2t.
\end{array}
\label{bolzmann}
\end{eqnarray}

The matrix elements for $\lambda$'s with all parts unequal ($\lambda_i\ne\lambda_{i+1}$ for all $i$) can be written as:
\begin{eqnarray}
\Lambda^s|\mu\rangle=\prod_1^M D_{\mu_{i}\lambda_i}\bar D_{\lambda_i\mu_{i-1}}|\lambda\rangle
\label{elements}
\end{eqnarray}
where $\mu_{0}=\mu_M+N$, and:
\begin{eqnarray}
  D_{mn}&=&
\begin{array}{ll}
    \omega_3^{n-m-1}\omega_6,\ m<n\\ 
    \omega_4/\omega_5,\ m=n
    \end{array}
     \nonumber\\
     \bar D_{nm}&=&
     \begin{array}{ll}
     \omega_1^{m-n-1}\omega_5,\ n<m\\ 
    \omega_2/\omega_6,\ n=m.
    \end{array}
\label{DDdef}
\end{eqnarray}
To properly evaluate 
the matrix elements $\Lambda^s_{\lambda\mu}$ when the partitions $\lambda$ has equal parts, 
we must correct the boundary terms $\lambda_i=\lambda_{i-1}=\mu_{i-1}$
by substituting $\omega_0\omega_5$ to $\omega_2\omega_4$ for each occurrence of 
$\bar D_{\mu_{i}\mu_{i}} D_{\mu_{i}\mu_{i}}$.

For the eigenvector, we take the ansatz:
\begin{eqnarray}
R_\mu=\sum_{P\in S_M} B(P)\xi_{P_1}^{\mu_1}\cdots\xi_{P_M}^{\mu_M}
\label{betheeigenvector}
\end{eqnarray}
where $B(P)$ are amplitudes to be determined, and we assume it is valid for $\mu_0\ge \mu_1\cdots\ge \mu_M$.

Let us write the (incorrect) eigenvalue equation as if (\ref{elements}) was still valid for $\lambda$'s with 
equal parts:
\begin{eqnarray}
(RA^s)_\mu=\sum_{P\in S_M} B(P)\sum_{\mu_i\leq\lambda_i\leq\mu_{i-1}}\prod_1^M D_{\mu_{i}\lambda_i}\bar D_{\lambda_i\mu_{i-1}}\xi_{P_i}^{\lambda_i}.
\label{betheeigenvector2}
\end{eqnarray}

For a fixed permutation, we can effect the sum over each $\lambda_i$ independently in (\ref{betheeigenvector2}), and using:
\begin{eqnarray}
\sum_{m_2\leq n\leq m_1}  D_{m_2n}\bar D_{nm_1}\xi^{n}&=&
({\xi\omega_5\omega_6\over \omega_1-\xi\omega_3}+\omega_4)\omega_1^{m_2-m_1-1}\xi^{m_2}+
({\omega_2}-{\omega_5\omega_6\over \omega_1-\xi\omega_3})\omega_3^{m_2-m_1-1}\xi^{m_1}\nonumber\\
&=&X_2+Y_1
\label{betheeigenvector21}
\end{eqnarray}
where the subscript of $X,Y$ coincides with that of the exponent of $\xi$,
we get for $P=1$:
\begin{eqnarray}
\sum_{\mu_i\leq\lambda_i\leq\mu_{i-1}}\prod_1^M D_{\mu_{i}\lambda_i}\bar D_{\lambda_i\mu_{i-1}}\xi_{i}^{\lambda_i}=(
{\contraction{X_M+}{Y}{_{M-1})(}{X}
\contraction{X_M+Y_{M-1})(X_{M-1}\cdots}{Y}{_1)(}{X} 
\contraction[2ex] {}{X}{_M+Y_{M-1})(X_{M-1}\cdots Y_1)(X_1+}{Y}
X_M+Y_{M-1})(X_{M-1}\cdots Y_1)(X_1+Y_0})
\label{betheeigenvector3}
\end{eqnarray}
where it is understood that $X_i+Y_{i-1}$ is a function of $\xi_i$.
To recover the correct eigenvalue equation, the term $\omega_2\omega_4$ in the products $\contraction{}{Y_i}{}{X_i}Y_iX_i$ needs to be replaced by
$\omega_0\omega_5$.

We can choose the amplitudes $B(P)$ such that after the summation over the permutations, 
only the terms $X_1\cdots X_M$, $Y_0\cdots Y_{M-1}$,
survive.  this will be the case if we can cancel the terms $Y_iX_i$ by requiring:
\begin{eqnarray}
B(P)
\contraction{}{Y}{_i(\xi_{P_{i+1}})}{X}
Y_i(\xi_{P_{i+1}})X_i(\xi_{P_i})
+B(P({i,i+1}))
\contraction{}{Y}{_i(\xi_{P_{i}})}{X}
Y_i(\xi_{P_{i}})X_i(\xi_{P_{i+1}})=0
\label{betheeigenvector4}
\end{eqnarray}
where $(i,j)$ denotes the permutation of $i$ with $j$.
The product $(\xi_{P_i}\xi_{P_{i+1}})^{\lambda_i}$ factors out, and
using the Orbach parametrization:
\begin{eqnarray}
\xi_i=\xi( u_i)={  u_i+s \over 1+ u_is}
\label{betheeigenvector5}
\end{eqnarray}
the condition (\ref{betheeigenvector4}) becomes:
\begin{eqnarray}
{B(P)\over B(P({i,i+1}))}=-{  u_{P_{i}}-t u_{P_{i+1}} \over{ u_{P_{i+1}}-t u_{P_{i}}}}
\label{betheeigenvector6}
\end{eqnarray}
with the solution $B(P)=B( u_{P_1},\cdots, u_{P_M})$:
\begin{eqnarray}
{B( u_k)}=\prod_{i<j} { u_{i}-t u_{{j}}\over { u_{i}- u_{{j}}}}
\label{betheeigenvector7}
\end{eqnarray}


In the closed chain case, we must take into account the periodicity conditions $S^+_{k+N}=XS^+_k,\ S^-_{k+N}=X^{-1}S^-_k$, which impose:
\begin{eqnarray}
R_{\mu_{n+N},\mu_1,\cdots\mu_{n-1}}=XR_{\mu_1,\cdots\mu_{n}},
\label{Rsper}
\end{eqnarray}
or:
\begin{eqnarray}
B(P)\xi_{P1}^N=XB(PC)
\label{Rsper1}
\end{eqnarray}
where $C_k=k+1$. Then, using (\ref{betheeigenvector6}) to express $B(PC)/B(P)$ as a product over the two-body scatterings, 
this condition gives the Bethe equations determining the $ u_i$'s:
\begin{eqnarray}
X\xi( u_k)^{-N}=\prod_{l\ne k}{ u_k-t u_l\over t u_k- u_l},
\label{bethef1}
\end{eqnarray}
and the eigenvalue is  the prefactor of $\prod\xi_{i}^{\lambda_i}$ in $X_1\cdots X_M+Y_0\cdots Y_{M-1}$
in (\ref{betheeigenvector3}) equal to:
\begin{eqnarray}
\tilde\Lambda_N^s (z U)=\omega_1^N\prod_{i=1}^M{1-zt u_i\over 1-z u_i}+X\omega_3^Nt^M\prod_{i=1}^M{1-z u_i/t\over 1-z u_i}.
\label{betheeigenvector10}
\end{eqnarray}

The preceding discussion only proves the validity of the ansatz (\ref{betheeigenvector}) with $B(P)$ given
by (\ref{betheeigenvector7}) for configurations $\lambda$ with $m_k(\lambda)\le 2$,
and we will assume it is correct  for $m_k(\lambda)$ arbitrary nonnegative integers.

\end{appendix}
\providecommand{\newblock}{}

\end{document}